\documentclass[aps,pra,showpacs,twocolumn,superscriptaddress]{revtex4}
\usepackage{amsmath}
\usepackage{amsfonts}
\usepackage{amssymb}
\usepackage{graphicx}
\usepackage{bbm}
\usepackage{natbib}
\usepackage[dvipdfm]{hyperref}
\makeatletter

\begin{document}

\title{Effect of bath temperature on the decoherence of quantum dissipative systems}

\author{Wei Wu}

\email{weiwu@csrc.ac.cn}

\affiliation{Beijing Computational Science Research Center, Beijing 100193, People's Republic of China}

\affiliation{Zhejiang Institute of Modern Physics and Physics Department, \\
Zhejiang University, Hangzhou 310027, China}

\author{Hai-Qing Lin}

\affiliation{Beijing Computational Science Research Center, Beijing 100193, People's Republic of China}

\begin{abstract}
We report an anomalous decoherence phenomenon of a quantum dissipative system in the framework of a stochastic decoupling scheme along with a hierarchical equations-of-motion formalism without the usual Born-Markov or weak coupling approximations. It is found that the decoherence of a two-qubit spin-boson model can be reduced by increasing the bath temperature in strong-coupling regimes. For the weak-coupling situation, we find that the bath temperature may enhance the decoherence. This result is contrary to the common recognition that a higher bath temperature always induces a more severe decoherence and suggests that a decoherence dynamical transition occurs in this two-qubit spin-boson model. We also demonstrate that the critical transition point can be characterized by the behavior of the frequency spectrum of the quantum coherence indicator.
\end{abstract}
\maketitle

\section{Introduction}\label{sec:sec1}

The rapid development of nanotechnology has opened the possibility to realize quantum information tasks at an atomic scale in experiment~\cite{1,2}. On the other hand, due to the unavoidable coupling with the surrounding bath, the microscopic quantum device severely undergoes decoherence which is the main difficulty in fulfilling reliable quantum computation and quantum communication tasks.

The dissipation-induced decoherence in a quantum microscopic system can be effectively modeled by a spin-boson model~\cite{3,4} which describes the interaction between a quantum system and a bosonic bath. The spin-boson model has attracted considerable attention in past decades because it provides a universal model for numerous physical and chemical processes. The reduced system dynamics of spin-boson model has been studied by various analytical and numerical methods, for example, the polaron~\cite{5,6} or generalized Silbey-Harris~\cite{7,8,9,10,11,add0} transformation approach, the time-dependent numerical renormalization group method~\cite{12,13,14,add1}, the quasi adiabatic propagator path integral,~\cite{9,15,16} and hierarchical equations of motion (HEOM)~\cite{17,18,19,20,21,22,23,add2}. Each method has its own regimes of validity depending on the system-bath coupling strength, the bath temperature and the bath spectral density function. HEOM is a set of time-local equations for the reduced system, which was originally proposed by Tanimura and his co-workers~\cite{17,18} as a nonperturbative numerical method. In recent years, HEOM was successfully used to study the dynamics of chemical and biophysical systems, such as optical line shapes of molecular aggregates~\cite{24} and electron energy transfer dynamics in the Fenna-Matthews-Olson complex~\cite{25,26}.

Thermal noise is one of the most familiar reasons for decoherence. It is a common recognition that the bosonic bath temperature always plays a negative role in preserving the coherence, i.e., a higher bosonic bath temperature induces a more severe decoherence~\cite{27,28,29}. However, it was pointed out in Refs.~\cite{11,30} that the bosonic bath temperature can enhance the coherence in a two-qubit spin-boson model, where a bare qubit interacts with the other qubit which is coupled to a thermal bosonic bath. Nevertheless, their conclusions are based on the Born-Markov or weak coupling approximations. A very interesting question arises: Does this interesting decoherence phenomenon (decoherence reduced by increasing bath temperature) still exist without any of these approximations? To answer this question, in this paper, we reexamine the decoherence dynamics of this two-qubit spin-boson model by making use of a stochastic decoupling scheme along with HEOM~\cite{31,32,33} which is beyond the usual Born-Markov or weak-coupling approximations. It is found that the decoherence can be reduced by increasing the bath temperature in strong-coupling regimes; for weak coupling, the bath temperature may enhance the decoherence. Our study is the generalization of previous studies~\cite{11,30} and suggests a decoherence dynamical transition in this two-qubit spin-boson model.

This paper is organized as follows. In Sec.~\ref{sec:sec2}, we briefly outline the formalism of stochastic decoupling scheme along with HEOM for quantum dissipative system. In Sec.~\ref{sec:sec3}, we adopt HEOM to study the decoherence dynamics of a two-qubit spin-boson model and compare the numerical results with those of Born-Markov master equation. Finally, some discussions and conclusions are drawn in Sec.~\ref{sec:sec4}.

\section{Formulation}\label{sec:sec2}

We start with a general quantum dissipative system whose Hamiltonian $\hat{H}$ can be described as follows:
\begin{equation}\label{eq:eq1}
\hat{H}=\hat{H}_{s}+\hat{H}_{b}+f(\hat{s})g(\hat{b}),
\end{equation}
where $\hat{H}_{s}$ is the Hamiltonian of the subsystem of interest and $f(\hat{s})$ denotes the subsystem's operator coupled to its surrounding bath. The Hamiltonian of the bath is $\hat{H}_{b}=\sum_{k}\omega_{k}\hat{a}_{k}^{\dagger}\hat{a}_{k}$, where $\hat{a}_{k}^{\dagger}$ and $\hat{a}_{k}$ are the creation and annihilation operators of the $k$th harmonic oscillators, respectively. The $g(\hat{b})$ represents the bath operator, and we assume $g(\hat{b})=\sum_{k}g_{k}(a_{k}^{\dagger}+a_{k})$ through this paper.

The complexity of a quantum dissipative system lies in the interaction between the subsystem and its surrounding bath which can be decoupled by making use of the approach proposed by Shao \emph{et al}.~\cite{31,32,33}. By the decoupling method, the dynamical evolution of the bath will be no longer involved in the dynamical evolution of the subsystem, which is very helpful to study the dissipative system dynamics. As a result, the density matrix of the whole system, $\hat{\rho}(t)$, can be expressed as~\cite{31,32,33}
\begin{equation}\label{eq:eq2}
\hat{\rho}(t)=\mathcal{M}\{\hat{\rho}_{s}(t)\hat{\rho}_{b}(t)\},
\end{equation}
where we have assumed the whole system is initially prepared in a product state $\hat{\rho}(0)=\hat{\rho}_{s}(0)\hat{\rho}_{b}(0)$ and $\mathcal{M}\{...\}$ is the ensemble mean operation over noises. The density matrices $\hat{\rho}_{s}(t)$ and $\hat{\rho}_{b}(t)$ obey the following stochastic differential equations, respectively~\cite{31,32,33}
\begin{equation}\label{eq:eq3}
\begin{split}
id\hat{\rho}_{s}(t)=&[\hat{H}_{s},\hat{\rho}_{s}(t)]dt+\frac{1}{2}[f(\hat{s}),\hat{\rho}_{s}(t)]d\varpi_{1t}\\
&+\frac{i}{2}\{f(\hat{s}),\hat{\rho}_{s}(t)\}d\varpi_{2t}^{\ast},
\end{split}
\end{equation}
and
\begin{equation}\label{eq:eq4}
\begin{split}
id\hat{\rho}_{b}(t)=&[\hat{H}_{b},\hat{\rho}_{b}(t)]dt+\frac{1}{2}[g(\hat{b}),\hat{\rho}_{b}(t)]d\varpi_{2t}\\
&+\frac{i}{2}\{g(\hat{b}),\hat{\rho}_{b}(t)\}d\varpi_{1t}^{\ast},
\end{split}
\end{equation}
where $d\varpi_{1t}=[\mu_{1}(t)+i\mu_{4}(t)]dt$ and $d\varpi_{2t}=[\mu_{2}(t)+i\mu_{3}(t)]dt$ are complex-valued Wiener processes, and $\mu_1(t)$ and $\mu_2(t)$ are two uncorrelated white noises which satisfy $\mathcal{M}\{\mu_{j}(t)\}=0$ and $\mathcal{M}\{\mu_{j}(t)\mu_{k}(t')\}=\delta_{jk}\delta(t-t')$. Here the commutation relations are defined as $[\hat{X},\hat{Y}]\equiv \hat{X}\hat{Y}-\hat{Y}\hat{X}$ and $\{\hat{X},\hat{Y}\}\equiv \hat{X}\hat{Y}+\hat{Y}\hat{X}$.

The reduced density matrix of the subsystem, $\tilde{\rho}_{s}(t)$, is defined by $\tilde{\rho}_{s}(t)\equiv tr_{b}[\hat{\rho}(t)]=\mathcal{M}\{\hat{\rho}_{s}tr_{b}[\hat{\rho}_{b}(t)]\}$ which contains all the physical information of the subsystem of interest and $tr_{b}[\hat{\rho}_{b}(t)]=\exp\{\int_{0}^{t}d\tau[\mu_{1}(\tau)-i\mu_{4}(\tau)]\bar{g}(\tau)\}$ with $\bar{g}(t)=tr_{b}[\hat{\rho}_{b}(t)g(\hat{b})]/tr_{b}[\hat{\rho}_{b}(t)]$. By employing a Girsanov transformation~\cite{31,32,33}, we can absorb $tr_{b}[\hat{\rho}_{b}(t)]$ into the measure of stochastic processes and obtain the stochastic equation of $\hat{\rho}_{s}(t)$ as follows:
\begin{equation}\label{eq:eq5}
\begin{split}
id\hat{\rho}_{s}(t)=&[\hat{H}_{s}+f(\hat{s})\bar{g}(t),\hat{\rho}_{s}(t)]dt+\frac{1}{2}[f(\hat{s}),\hat{\rho}_{s}(t)]d\varpi_{1t}\\
&+\frac{i}{2}\{f(\hat{s}),\hat{\rho}_{s}(t)\}d\varpi_{2t}^{\ast}.
\end{split}
\end{equation}
From the stochastic equation above, it is clear to see that $\bar{g}(t)$ plays a similar role to that of the influence functional in the path integral treatment~\cite{31,34}, and $\bar{g}(t)$ is the bath induced mean field which fully characterizes the influence of the bath on the subsystem. Solving the evolution equation of the bath, one can obtain the expression of the bath-induced mean field as follows:
\begin{equation}\label{eq:eq6}
\begin{split}
\bar{g}(t)=&\int_{0}^{t}C_{R}(t-\tau)d\omega_{1\tau}^{*}+C_{I}(t-\tau)d\omega_{2\tau}\\
=&\bar{g}_{1}(t)+\bar{g}_{2}(t),
\end{split}
\end{equation}
where
\begin{equation*}
\bar{g}_{1}(t)=\frac{1}{2}\int_{0}^{t}d\tau C(t-\tau)[\mu_{1}(\tau)-i\mu_{4}(\tau)-i\mu_{2}(\tau)+\mu_{3}(\tau)],
\end{equation*}
\begin{equation*}
\bar{g}_{2}(t)=\frac{1}{2}\int_{0}^{t}d\tau C^{*}(t-\tau)[\mu_{1}(\tau)-i\mu_{4}(\tau)+i\mu_{2}(\tau)-\mu_{3}(\tau)],
\end{equation*}
with $C_{R}(t)$ and $C_{I}(t)$ being the real and imaginary parts of the bath correlation function $C(t)$, respectively. Assuming the bath is in a thermal equilibrium state $\hat{\rho}_{b}(0)=\hat{\rho}_{th}=e^{-\hat{H}_{b}T^{-1}}/tr_{b}(e^{-\hat{H}_{b}T^{-1}})$ with the Boltzmann constant $k_{B}=1$, then one can obtain
\begin{equation}\label{eq:eq7}
C(t)=\int_{0}^{\infty}{d\omega}J(\omega)[\coth(\frac{\omega}{2T})\cos(\omega{t})-i\sin(\omega{t})],
\end{equation}
where $J(\omega)=\sum_{k}g_{k}^{2}\delta(\omega-\omega_{k})$ is the bath spectral density function. Taking the ensemble mean operation on both sides of Eq.~\ref{eq:eq5}, one can finally obtain the motion equation of $\tilde{\rho}_{s}(t)$ as follows~\cite{31,32,33}
\begin{equation}\label{eq:eq8}
i\frac{d}{dt}\tilde{\rho}_{s}(t)=[\hat{H}_{s},\tilde{\rho}_{s}(t)]+[f(\hat{s}),\mathcal{M}\{\bar{g}_{1}(t)\hat{\rho}_{s}(t)+\bar{g}_{2}(t)\hat{\rho}_{s}(t)\}].
\end{equation}

Eq.~\ref{eq:eq8} is an exact motion equation for the reduced density matrix $\tilde{\rho}_{s}(t)$, though the general relation between $\tilde{\rho}_{s}(t)$ and $\mathcal{M}\{\bar{g}_{1,2}(t)\hat{\rho}_{s}(t)\}$ is unknown which is also the main difficulty in solving Eq.~\ref{eq:eq8}, because the stochastic simulation of $\mathcal{M}\{\bar{g}_{1,2}(t)\hat{\rho}_{s}(t)\}$ is not very effective, especially for studying the long-time effects. However, if the bath correlation function $C(t)$ can be written as a sum of exponentials~\cite{35,36,37}, this problem can be solved by making use of HEOM. First, we consider the simplest case,
\begin{equation}\label{eq:eq9}
C(t)=\alpha e^{-\beta t},
\end{equation}
where $\alpha$ and $\beta$ are assumed to be complex numbers for generality. For such an exponential bath correlation function, it is easy to find
\begin{equation*}
\frac{d}{dt}\bar{g}_{1}(t)=-\beta\bar{g}_{1}(t)+\frac{1}{2}\alpha[\mu_{1}(t)-i\mu_{4}(t)-i\mu_{2}(t)+\mu_{3}(t)],
\end{equation*}
\begin{equation*}
\frac{d}{dt}\bar{g}_{2}(t)=-\beta^{*}\bar{g}_{1}(t)+\frac{1}{2}\alpha^{*}[\mu_{1}(t)-i\mu_{4}(t)+i\mu_{2}(t)-\mu_{3}(t)].
\end{equation*}
Thus, one can obtain
\begin{equation*}
\begin{split}
i\frac{d}{dt}\tilde{\rho}_{10}(t)=&-i\beta\tilde{\rho}_{10}(t)+\alpha f(\hat{s})\tilde{\rho}_{00}(t)+[\hat{H}_{s}, \tilde{\rho}_{10}(t)]\nonumber\\
&+[f(\hat{s}),\tilde{\rho}_{20}(t)]+[f(\hat{s}),\tilde{\rho}_{11}(t)],
\end{split}
\end{equation*}
\begin{equation*}
\begin{split}
i\frac{d}{dt}\tilde{\rho}_{01}(t)=&-i\beta^{*}\tilde{\rho}_{01}(t)-\alpha^{*}\tilde{\rho}_{00}(t)f(\hat{s})\nonumber\\
&+[\hat{H}_{s}, \tilde{\rho}_{01}(t)]+[f(\hat{s}),\tilde{\rho}_{11}(t)]+[f(\hat{s}),\tilde{\rho}_{02}(t)],
\end{split}
\end{equation*}
where we have defined the auxiliary matrices $\tilde{\rho}_{mn}(t)\equiv \mathcal{M}\{\bar{g}_{1}^{m}\bar{g}_{2}^{n}\hat{\rho}_{s}(t)\}$ with $\tilde{\rho}_{00}(t)=\tilde{\rho}_{s}(t)$. These two equations are not closed because they are coupled to more unknown terms $\tilde{\rho}_{20}(t)$, $\tilde{\rho}_{11}(t)$, and $\tilde{\rho}_{01}(t)$. However, we can repeat the above procedure, i.e., taking the time derivatives of $\tilde{\rho}_{20}(t)$, $\tilde{\rho}_{11}(t)$, and $\tilde{\rho}_{01}(t)$, and finally obtain a set of coupled ordinary differential equations as follows
\begin{equation}\label{eq:eq10}
\begin{split}
\frac{d}{dt}\tilde{\rho}_{\vec{l}}(t)=&(-i\hat{H}_{s}^{\times}-\vec{l}\cdot\vec{\beta})\tilde{\rho}_{\vec{l}}(t)+\hat{\Phi}\sum_{p=1}^{2}\tilde{\rho}_{\vec{l}+\vec{e}_{p}}(t)\\
&+\sum_{p=1}^{2}l_{p}\hat{\Psi}_{p}\tilde{\rho}_{\vec{l}-\vec{e}_{p}}(t),
\end{split}
\end{equation}
where $\vec{l}=(m,n)$, $\vec{\alpha}=(\alpha,\alpha^{*})$, $\vec{\beta}=(\beta,\beta^{*})$, $\vec{e}_{1}=(1,0)$, $\vec{e}_{2}=(0,1)$,
\begin{equation*}
\hat{\Phi}=-if(\hat{s})^{\times},~~~\hat{\Psi}_{p}=\frac{i}{2}\alpha_{p}[(-1)^{p}f(s)^{\circ}-f(s)^{\times}],
\end{equation*}
and we have introduced two superoperators, $\hat{X}^{\times}\hat{Y}\equiv [\hat{X},\hat{Y}]$ and $\hat{X}^{\circ}\hat{Y}\equiv \{\hat{X},\hat{Y}\}$. The initial state conditions of these auxiliary matrices are $\tilde{\rho}_{00}(0)=\tilde{\rho}_{s}(0)$ and $\tilde{\rho}_{\vec{l}\neq(0,0)}(0)=0$. For numerical simulations, we need to truncate the number of HEOM for a sufficiently large integer $N$, which means all terms $\tilde{\rho}_{mn}(t)$ with $m+n>N$ are set to zero and forms a closed set of differential equations. We can increase the hierarchy order $N$ until the result of $\tilde{\rho}_{s}(t)$ converges. In this sense, we convert the stochastic differential equation of Eq.~\ref{eq:eq8} into a set of ordinary differential equations which are convenient for numerical simulation. A similar HEOM can be also derived by employing the superoperator technique~\cite{22} or Feynman-Vernon influence functional approach~\cite{23}.

\begin{figure}
\centering
\includegraphics[angle=0,width=7cm]{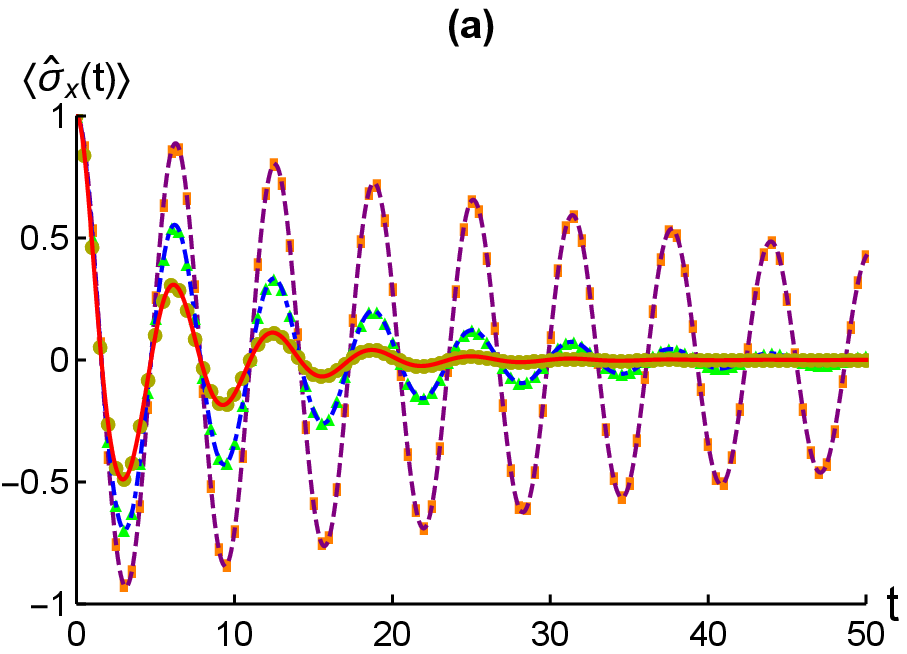}
\includegraphics[angle=0,width=7cm]{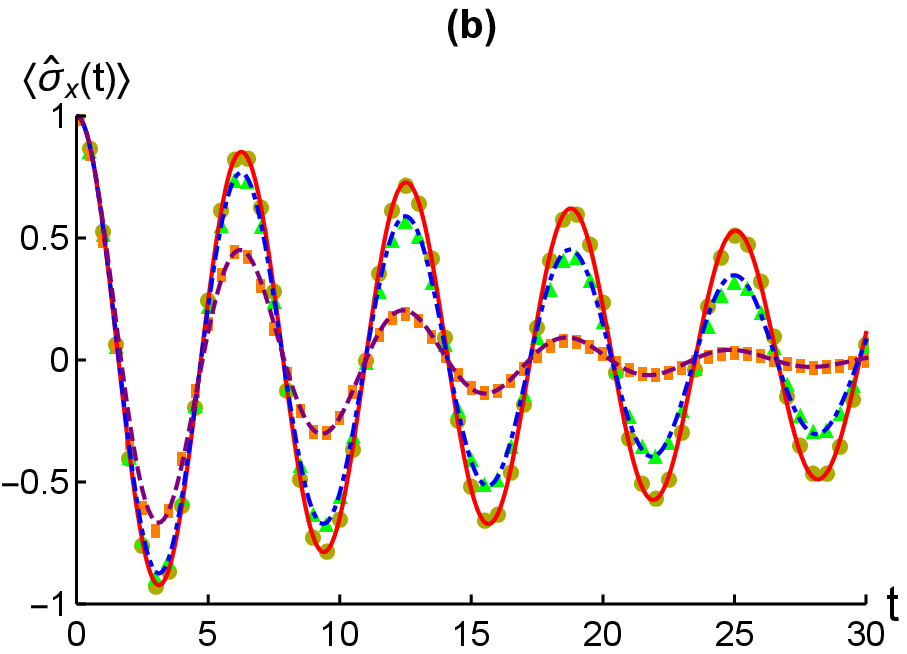}
\caption{\label{fig:fig1} (a) The quantum coherence indicator $\langle\hat{\sigma}_{x}(t)\rangle$ versus $t$ at zero temperature with different coupling parameters: $\lambda=0.01\omega_{0}$ (numerical results, purple dashed line; analytical results, orange rectangles), $\lambda=0.05\omega_{0}$ (numerical results, blue dotdashed line; analytical results, green triangles) and $\lambda=0.1\omega_{0}$ (numerical results, red solid line; analytical results, yellow circles). Other parameters are chosen as $\gamma=0.5\omega_{0}$ and $\omega_{0}=1$. (b) The quantum coherence indicator $\langle\hat{\sigma}_{x}(t)\rangle$ versus $t$ at different temperatures: $T^{-1}=0.01\omega_{0}^{-1}$ (numerical results, purple dashed line; analytical results, orange rectangles), $T^{-1}=0.03\omega_{0}^{-1}$ (numerical results, blue dotdashed line; analytical results, green triangles) and $T^{-1}=0.05\omega_{0}^{-1}$ (numerical results: red solid line, analytical results: yellow circles). Other parameters are chosen as $\eta=5\times10^{-4}\omega_{0}$, $\omega_{c}=3\omega_{0}$, and $\omega_{0}=1$.}
\end{figure}

We now generalize this approach to the situation where the bath correlation function can be expressed as a sum of exponential functions, i.e.,
\begin{equation}\label{eq:eq11}
C(t)=\sum_{k=1}^{\epsilon}C_{k}(t)=\sum_{k=1}^{\epsilon}\alpha_{k}e^{-\beta_{k}t}.
\end{equation}
The sums of such exponentials are well suited to approximately describe the bath spectral density function at finite temperature and can be achieved for realistic application~\cite{35,36,37}. By making use of the same procedure outlined above, one can derive the following HEOM
\begin{equation}\label{eq:eq12}
\begin{split}
\frac{d}{dt}\tilde{\rho}_{\vec{\ell}}(t)=&(-i\hat{H}_{s}^{\times}-\vec{\ell}\cdot\vec{\beta})\tilde{\rho}_{\vec{\ell}}(t)+\hat{\Phi}\sum_{q=1}^{2\epsilon}\tilde{\rho}_{\vec{\ell}+\vec{e}_{q}}(t)\\
&+\sum_{q=1}^{2\epsilon}\ell_{q}\hat{\Psi}_{q}\tilde{\rho}_{\vec{\ell}-\vec{e}_{q}}(t),
\end{split}
\end{equation}
where $\vec{\ell}=(m_{1},n_{1},m_{2},n_{2},...,m_{\epsilon},n_{\epsilon})$, $\vec{\alpha}=(\alpha_{1},\alpha_{1}^{*},\alpha_{2},\alpha_{2}^{*},...,\alpha_{\epsilon},\alpha_{\epsilon}^{*})$, $\vec{\beta}=(\beta_{1},\beta_{1}^{*},\beta_{2},\beta_{2}^{*},...,\beta_{\epsilon},\beta_{\varepsilon}^{*})$, $\vec{e}_{q}=(0,0,0,...,1_{q},...0)$, and $\hat{\Psi}_{l}=\frac{i}{2}\alpha_{q}[(-1)^{q}f(\hat{s})^{\circ}-f(\hat{s})^{\times}]$. The corresponding auxiliary matrices are defined by
\begin{equation*}
\tilde{\rho}_{\vec{\ell}}(t)\equiv \mathcal{M}\{\prod_{k=1}^{\epsilon}\bar{g}_{k,1}^{m_{k}}\bar{g}_{k,2}^{n_{k}}\hat{\rho}_{s}(t)\},
\end{equation*}
where
\begin{equation*}
\bar{g}_{k,1}(t)\equiv \frac{1}{2}\int_{0}^{t}d\tau C_{k}(t-\tau)[\mu_{1}(\tau)-i\mu_{4}(\tau)-i\mu_{2}(\tau)+\mu_{3}(\tau)],
\end{equation*}
\begin{equation*}
\bar{g}_{k,2}(t)\equiv \frac{1}{2}\int_{0}^{t}d\tau C_{k}^{*}(t-\tau)[\mu_{1}(\tau)-i\mu_{4}(\tau)+i\mu_{2}(\tau)-\mu_{3}(\tau)].
\end{equation*}
It is necessary to point out that we did not use the usual Born-Markov or weak coupling approximations during the derivation of the HEOM and the result obtained by HEOM can be regarded as a rigorous numerical result.

\section{Results}\label{sec:sec3}

Now, we investigate the decoherence dynamics of a quantum dissipative system by employing the stochastic decoupling along with HEOM. But first, in order to verify the feasibility of this numerical scheme, we would like to compare the numerical results with the analytical results of a pure dephasing model~\cite{28}, where $\hat{H}_{s}=\frac{1}{2}\omega_{0}\hat{\sigma}_{z}$ and $f(\hat{s})=\hat{\sigma}_{z}$.
The reduced density matrix of this dephasing system can be exactly solved, and one can find the diagonal terms do not evolve in time, i.e., $\rho_{ee}(t)=\rho_{ee}(0)$ and $\rho_{gg}(t)=\rho_{gg}(0)$. The analytical expressions of nondiagonal terms are~\cite{28,38} $\rho_{eg}(t)=\rho_{ge}^{*}(t)=\rho_{eg}(0)\exp[-\Gamma(t)-i\omega_{0}t]$, where
\begin{equation}\label{eq:eq13}
\Gamma(t)=4\int_{0}^{\infty} d\omega J(\omega)\frac{1-\cos(\omega t)}{\omega^{2}}\coth(\frac{\omega}{2T}),
\end{equation}
is the decoherence factor. Then, it is easy to obtain the analytical expression of the physical quantity $\langle\hat{\sigma}_{x}(t)\rangle$ for initial state $\hat{\rho}_{s}(0)=\frac{1}{2}(|e\rangle\langle e|+|g\rangle\langle e|+|g\rangle\langle e|+|g\rangle\langle g|)$, where $|e,g\rangle$ are the eigenvalues of the Pauli $z$ matrix $\hat{\sigma}_{z}$, as follows
\begin{equation}\label{eq:eq14}
\langle\hat{\sigma}_{x}(t)\rangle\equiv tr[\tilde{\rho}_{s}(t)\hat{\sigma}_{x}]=\cos(\omega_{0}t)e^{-\Gamma(t)}.
\end{equation}
The oscillation amplitude of $\langle\hat{\sigma}_{x}(t)\rangle$ reflects the intensity of coherence in the quantum dissipative system~\cite{39,40}. In this paper, we choose $\langle\hat{\sigma}_{x}(t)\rangle$ as the quantum coherence indicator.

For the zero-temperature case, we assume the bath spectral density function $J(\omega)$ has a form of Lorentz spectrum type~\cite{22,39}
\begin{equation}\label{eq:eq15}
J(\omega)=\frac{1}{\pi}\frac{\lambda\gamma}{(\omega-\omega_{0})^{2}+\gamma^{2}},
\end{equation}
where $\lambda$ reflects the coupling strength between qubit and bath, and $\gamma$ is the broadening width of the bath mode which is connected to the bath correlation time, $\tau_{b}\sim\gamma^{-1}$. In this case, the bath correlation function is given by $C(t)=\lambda\exp[-(\gamma+i\omega_{0})t]$, which satisfies the condition to perform the HEOM scheme. In Fig.~\ref{fig:fig1}(a), we display the decoherence dynamics of $\langle\hat{\sigma}_{x}(t)\rangle$ obtained by numerical method as well as the analytical expression. It is clear to see that the numerical results are in good agreement with the analytical results regardless of weak-coupling or strong-coupling regimes.

\begin{figure}
\centering
\includegraphics[angle=0,width=4.2cm]{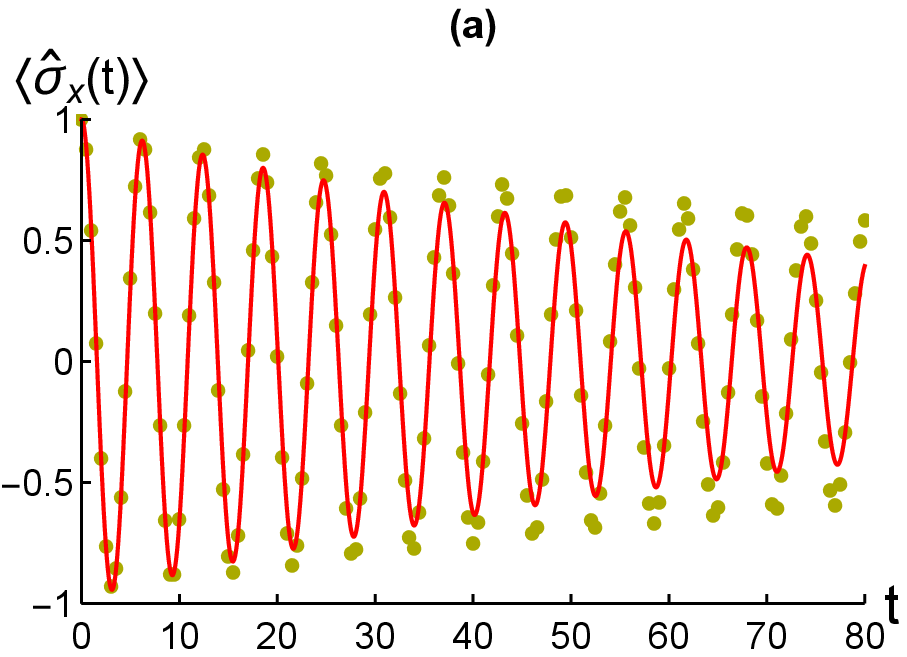}
\includegraphics[angle=0,width=4.2cm]{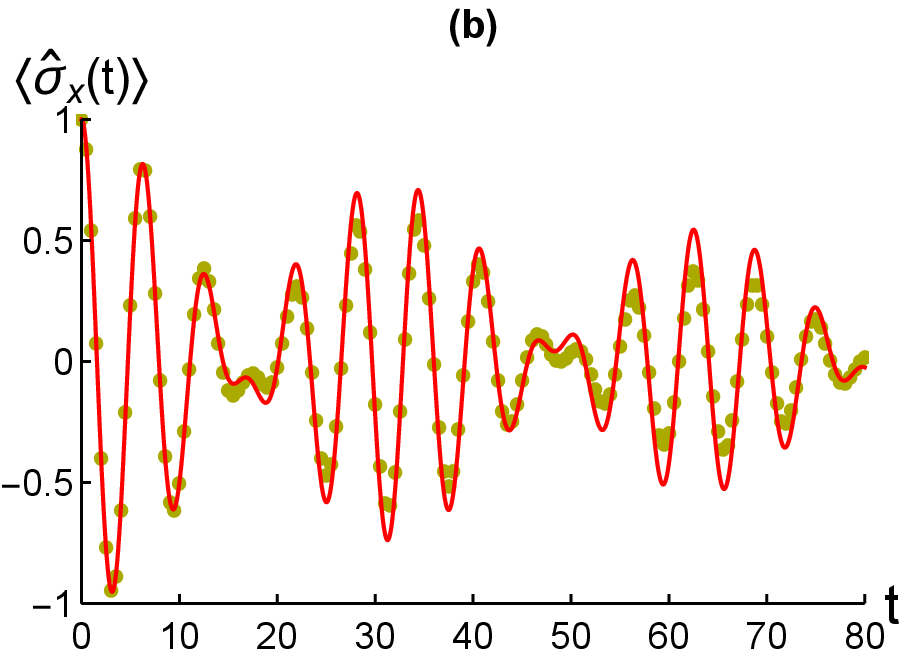}
\includegraphics[angle=0,width=4.2cm]{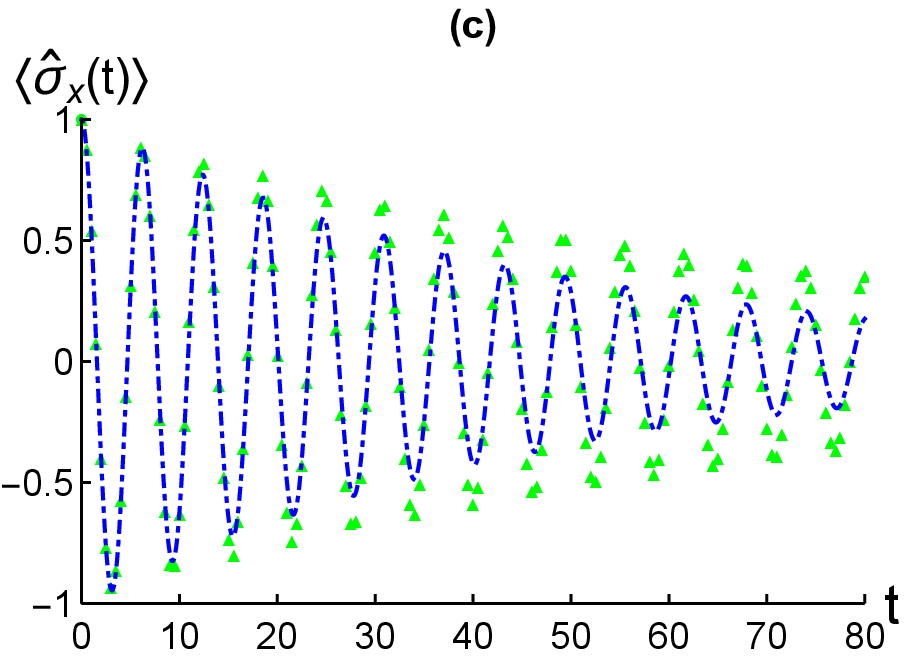}
\includegraphics[angle=0,width=4.2cm]{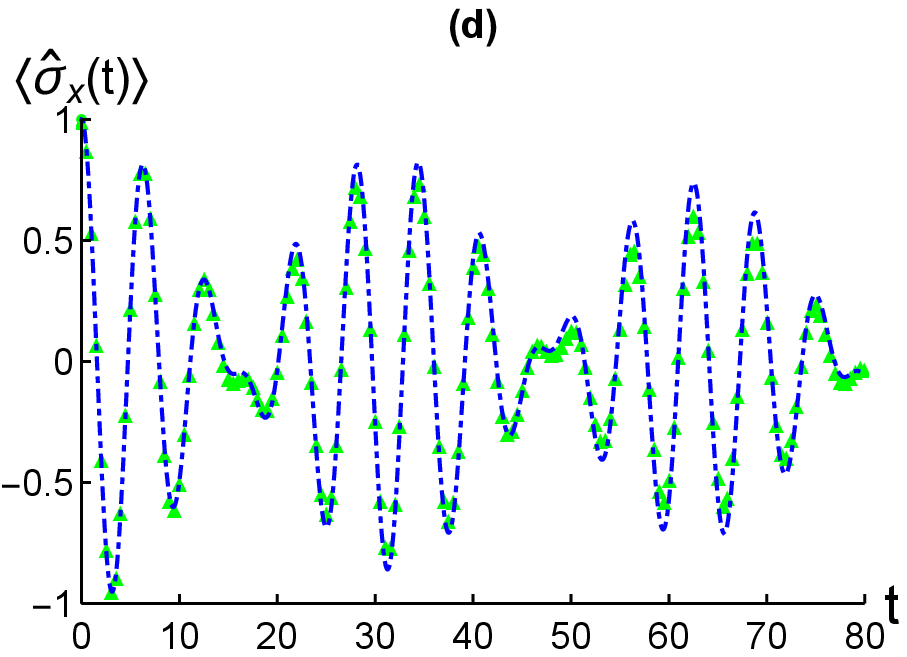}
\includegraphics[angle=0,width=4.2cm]{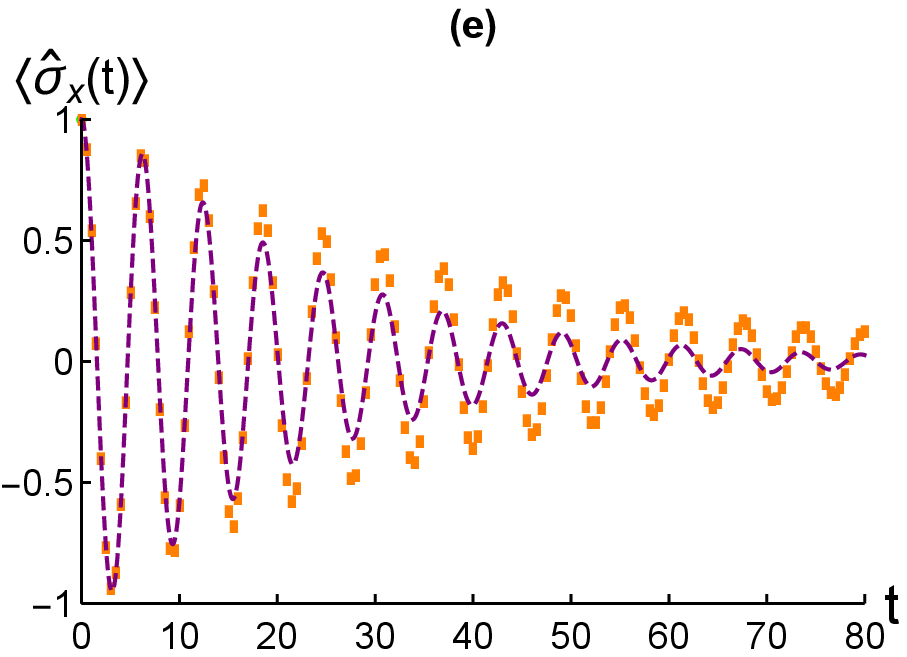}
\includegraphics[angle=0,width=4.2cm]{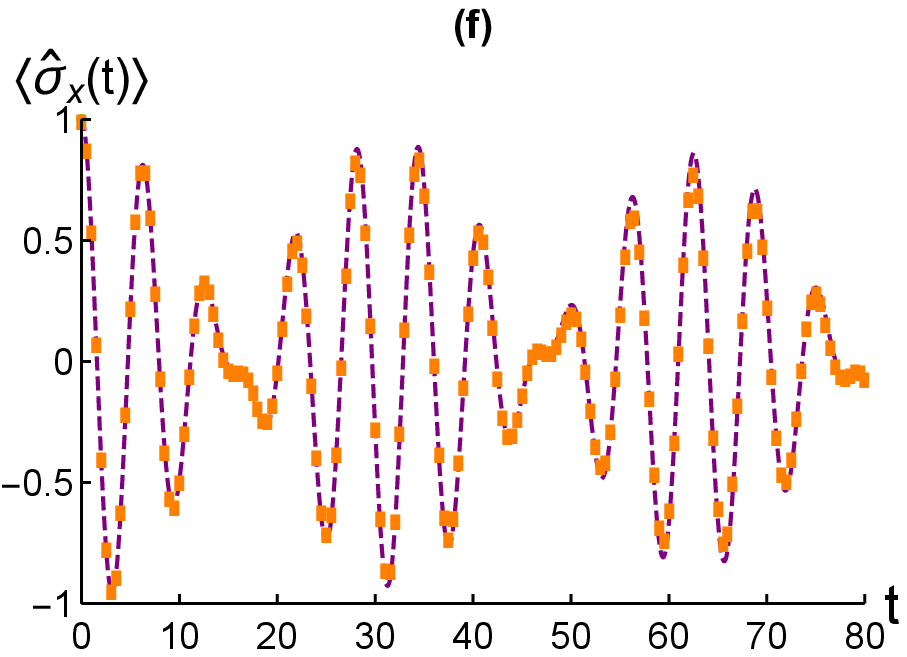}
\caption{\label{fig:fig2} The quantum coherence indicator $\langle\hat{\sigma}_{x}(t)\rangle$ of qubit $A$ versus $t$ at different temperatures in the strong-coupling regime $\eta=0.05\omega_{0}$: (a) $T^{-1}=0.05\omega_{0}^{-1}$ (numerical results, left red solid line; Born-Markov results, left yellow circles), (c) $T^{-1}=0.10\omega_{0}^{-1}$ (numerical results: left blue dotdashed line, Born-Markov results: left green triangles) and (e) $T^{-1}=0.20\omega_{0}^{-1}$ (numerical results, left purple dashed line; Born-Markov results, left orange rectangles). The $\langle\hat{\sigma}_{x}(t)\rangle$ of qubit $A$ versus $t$ at different temperatures in the weak coupling regime $\eta=0.001\omega_{0}$: (b) $T^{-1}=0.05\omega_{0}^{-1}$ (numerical results, right red solid line; Born-Markov results, right yellow circles), (d) $T^{-1}=0.10\omega_{0}^{-1}$ (numerical results, right blue dotdashed line; Born-Markov results, right green triangles) and (f) $T^{-1}=0.20\omega_{0}^{-1}$ (numerical results, left purple dashed line; Born-Markov results, left orange rectangles). Other parameters are chosen as $\omega_{c}=5\omega_{0}$ $g_{0}=0.1\omega_{0}$ and $\omega_{0}=1$.}
\end{figure}

For the finite-temperature case, we assume the bath density spectral function $J(\omega)$ is Ohmic spectrum with Drude cutoff throughout this paper:
\begin{equation}\label{eq:eq16}
J(\omega)=\frac{1}{\pi}\frac{2\eta\omega_{c}\omega}{\omega^{2}+\omega_{c}^{2}},
\end{equation}
where $\eta$ stands for the coupling strength between the subsystem and bath and $\omega_{c}$ is the cutoff frequency. In this case, the bath correlation function $C(t)$ is given by~\cite{17,18,19,20,21,22,25,26,32}
\begin{equation}\label{eq:eq17}
C(t)=\sum_{k=1}^{\infty}\alpha_{k}e^{-\beta_{k}t},
\end{equation}
where
\begin{equation*}
\alpha_{1}=\eta\omega_{c}\cot(\frac{\omega_{c}}{2T})-i\eta\omega_{c},~~~\beta_{1}=\omega_{c},
\end{equation*}
\begin{equation*}
\alpha_{k\geq 2}=4\eta\omega_{c}T\frac{\upsilon_{k}}{\upsilon_{k}^{2}-\omega_{c}^{2}},~~~\beta_{k\geq 2}=\upsilon_{k},
\end{equation*}
and $\upsilon_{k}\equiv2(k-1)\pi T$ denote the Matsubara frequencies. For numerical simulations, the bath correlation function can be approximately expressed as the sum of the first few terms in the series, this approximation is reliable when the bath temperature is not very low. In this paper, we add the number of Matsubara frequencies for a given bath temperature $T$ till the result converges. In Fig.~\ref{fig:fig1}(b), we compare the numerical results with the analytical results obtained by Eq.~\ref{eq:eq14}. It is clear to see numerical results coincide with exact analytical expression. These results convince us that this numerical scheme is reliable.

Next, we consider a two-qubit spin-boson model: a bare qubit interacts with the other one which is coupled to a thermal bath without rotating-wave approximations. The subsystem's Hamiltonian is given by
\begin{equation}\label{eq:eq18}
\hat{H}_{s}=\frac{1}{2}(\omega_{A}\hat{\sigma}_{z}^{A}\otimes\mathbf{\hat{1}}_{2}^{B}+\omega_{B}\mathbf{\hat{1}}_{2}^{A}\otimes\hat{\sigma}_{z}^{B})+g_{0}\hat{\sigma}_{x}^{A}\hat{\sigma}_{x}^{B},
\end{equation}
and $f(\hat{s})=\mathbf{\hat{1}}_{2}^{A}\otimes\hat{\sigma}_{x}^{B}$, where $\mathbf{\hat{1}}_{2}$ denotes a $2\times 2$ identity matrix. The parameter $g_{0}$ stands for the interaction strength between the two qubits. In this paper, we focus on the on-resonance case: $\omega_{A}=\omega_{B}=\omega_{0}$. This model has been studied in several previous articles~\cite{11,30} and has no exact analytical expression of the reduced density matrix for qubits. References~\cite{11,30} showed that bath temperature can enhance the coherence of qubit $A$; however, their conclusions are based on a Born-Markov approximation or weak-coupling approximation. In this paper, we recheck this conclusion by making use of stochastic decoupling along with HEOM which is beyond the usual Born-Markov and weak-coupling approximations.

In order to compare with the numerical method, we also derived the second-order Born-Markov master equation for Hamiltonian $\hat{H}=\hat{H}_{s}+\hat{H}_{b}+f(\hat{s})g(\hat{b})$ and the result is given by (see Appendix for details)
\begin{equation}\label{eq:eq19}
\frac{d}{dt}\tilde{\rho}_{s}(t)=[-i\hat{H}_{s}^{\times}-f(\hat{s})^{\times}\Upsilon(\hat{s})^{\times}+f(\hat{s})^{\times}\Xi(\hat{s})^{\circ}]\tilde{\rho}_{s}(t),
\end{equation}
where
\begin{equation*}
\Upsilon(\hat{s})\equiv\int_{0}^{\infty}d\tau C_{R}(\tau)\hat{f}_{s}(-\tau),
\end{equation*}
\begin{equation*}
\Xi(\hat{s})\equiv-i\int_{0}^{\infty}d\tau C_{I}(\tau)\hat{f}_{s}(-\tau),
\end{equation*}
with $\hat{f}_{s}(t)\equiv e^{i\hat{H}_{s}t}f(\hat{s})e^{-i\hat{H}_{s}t}$. If $[f(\hat{s}),\hat{H}_{s}]=0$, the above master equation reduces to the well-known Lindblad-type master equation~\cite{39,40}. In the case $[f(\hat{s}),\hat{H}_{s}]\neq0$, the interaction picture operator $\hat{f}_{s}(t)$ gains the difficulty in solving the Born-Markov master equation; however, by numerically diagonalizing $\hat{H}_{s}$, we can obtain two simpler expressions of $\Upsilon(\hat{s})$ and $\Xi(\hat{s})$, which will be very helpful to our calculation (see Appendix for details).

We plot the coherence indicator $\langle\hat{\sigma}_{x}(t)\rangle$ of qubit $A$ for the initial state
\begin{equation}\label{eq:eq20}
\hat{\rho}_{AB}(0)=\frac{1}{4}\left(
                                                                                      \begin{array}{cc}
                                                                                          1 & 1 \\
                                                                                          1 & 1 \\
                                                                                        \end{array}
                                                                                      \right)_{A}\otimes\left(
                                                                                        \begin{array}{cc}
                                                                                          1 & 1 \\
                                                                                          1 & 1 \\
                                                                                        \end{array}
                                                                                      \right)_{B}
\end{equation}
at different bath temperatures in Fig.~\ref{fig:fig2}. It is found that the coherence indicator $\langle\hat{\sigma}_{x}(t)\rangle$ exhibits a simple oscillation in strong coupling regimes and the oscillation amplitude becomes large with the increase of bath temperature $T$ regardless of Markovian or non-Markovian cases. This result coincides with previous studies~\cite{11,30}. However, with the decrease of system-bath coupling constant $\eta$, we find that the coherence indicator $\langle\hat{\sigma}_{x}(t)\rangle$ displays a collapse and revival phenomenon which results from the interference between two oscillations with different frequencies. In this situation, the quantum coherence of qubit $A$ is not meliorated but rather is damaged with the increase of bath temperature $T$.

This result suggests that there exists a critical coupling strength $\eta_{c}$ which represents the critical point from quantum beat dynamics to damped oscillation. If $\eta>\eta_{c}$, the coherence indicator $\langle\hat{\sigma}_{x}(t)\rangle$ exhibits a damped oscillation and the bath temperature can reduce the decoherence. While, if $\eta<\eta_{c}$, the coherence indicator $\langle\hat{\sigma}_{x}(t)\rangle$ displays a quantum beat dynamics~\cite{43,44} and the bath temperature enhances the decoherence.

Comparing with previous literature~\cite{11,30}, we reexamine the decoherence dynamics of the two-qubit spin-boson model by making use of the nonperturbative HEOM formalism. Our results demonstrate that the anomalous decoherence phenomenon (decoherence reduced by increasing bath temperature) is independent of the Born-Markov approximation. What is more important, we find the effects of bath temperature on the decoherence dynamics of this two-qubit spin-boson model are completely different for weak-coupling ($\eta<\eta_{c}$) and strong-coupling ($\eta>\eta_{c}$) cases. This result cannot be described by the usual weak coupling theory (say, the Lindblad master equation approach used in Ref.~\cite{30}) and shows that our nonperturbative formalism is able to extract more physical information about the decoherence of a quantum dissipative system. For more realistic quantum devices, it is important to realize that the effects of bath temperature on decoherence can be very intricate at a microscopic scale: the decoherence rate could not be a monotonic decreasing function of bath temperature.

\section{Discussion and conclusion}\label{sec:sec4}

\begin{figure}
\centering
\includegraphics[angle=0,width=4.2cm]{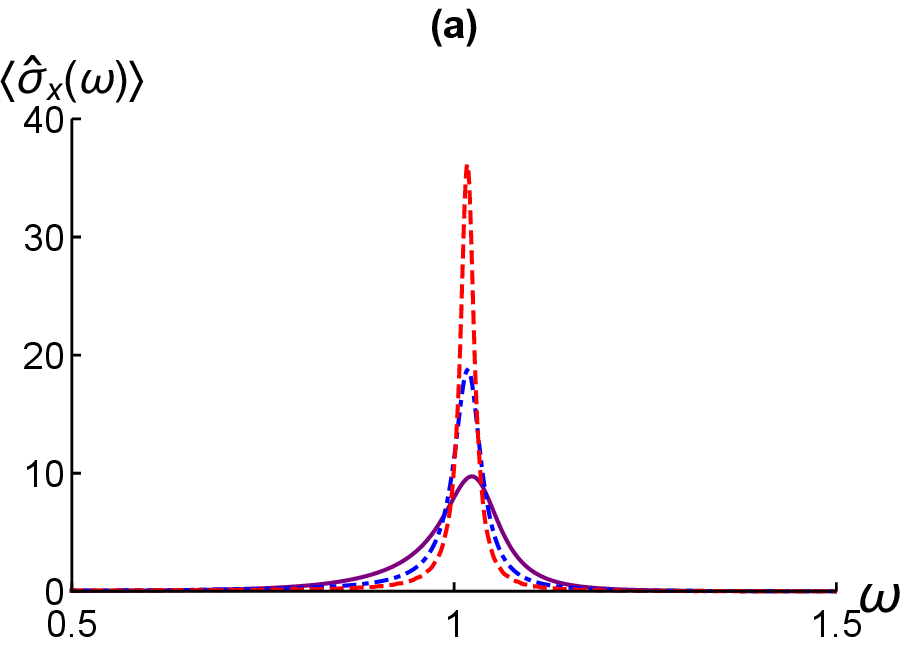}
\includegraphics[angle=0,width=4.2cm]{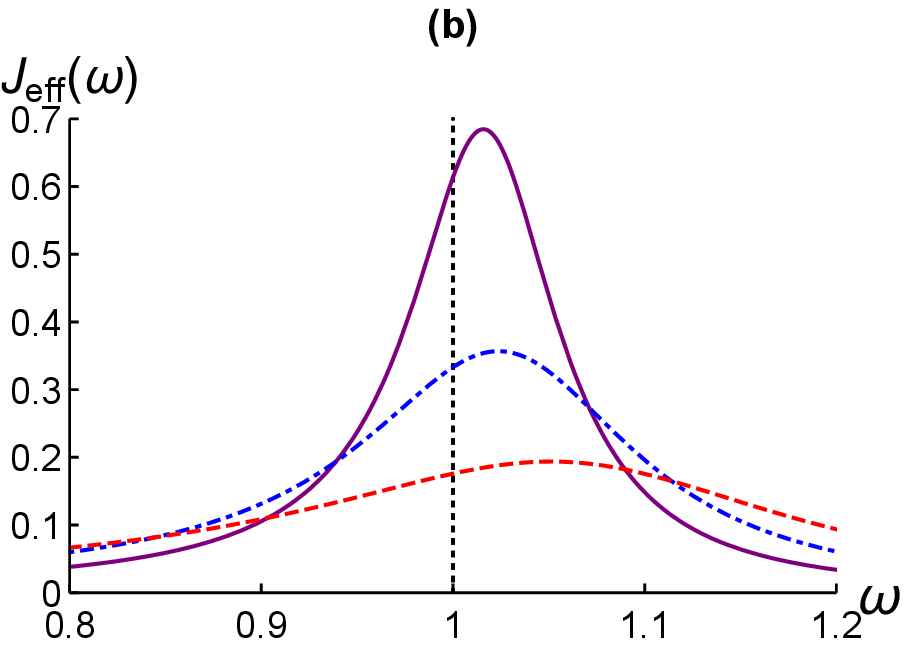}
\includegraphics[angle=0,width=4.2cm]{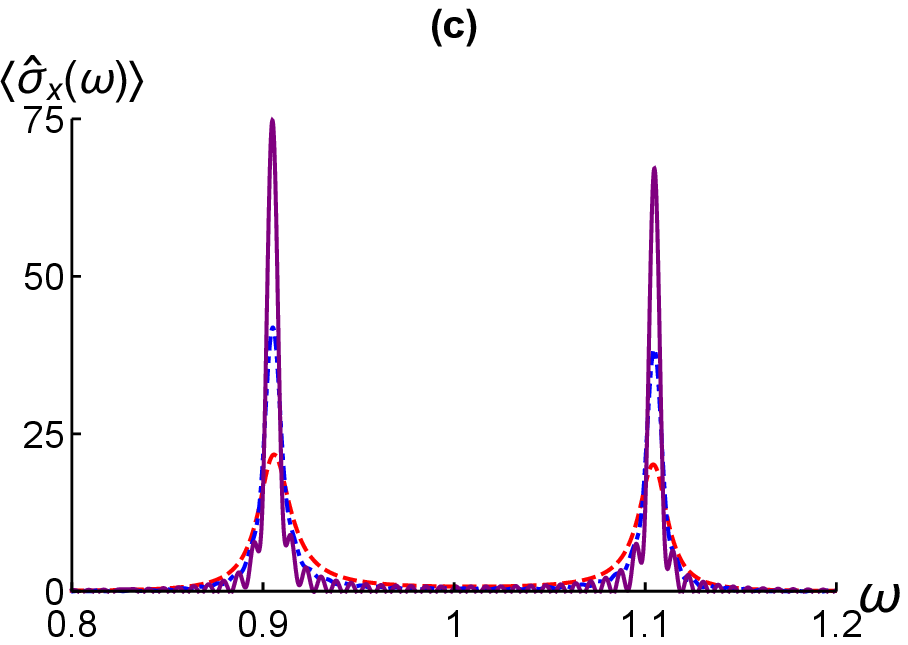}
\includegraphics[angle=0,width=4.2cm]{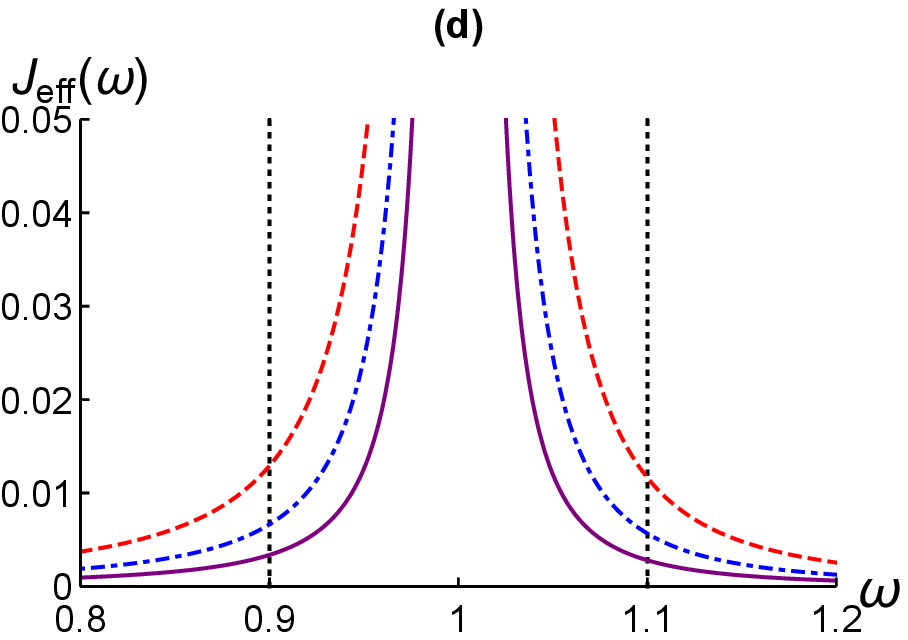}
\caption{\label{fig:fig3} The frequency spectrum $\langle\hat{\sigma}_{x}(\omega)\rangle$ of qubit $A$ versus $\omega$ in (a) strong-coupling regime $\eta=0.05\omega_{0}$ and (c) weak-coupling regime $\eta=0.001\omega_{0}$ with different bath temperatures: $T^{-1}=0.05\omega_{0}^{-1}$ (red dashed line), $T^{-1}=0.10\omega_{0}^{-1}$ (blue dot-dashed line), and $T^{-1}=0.20\omega_{0}^{-1}$ (purple solid line). The effective bath spectral density function $J_{\rm eff}(\omega)$ in (b) strong-coupling regime $\eta=0.05\omega_{0}$ and (d) weak-coupling regime $\eta=0.001\omega_{0}$ with different bath temperatures: $T^{-1}=0.05\omega_{0}^{-1}$ (red dashed line), $T^{-1}=0.10\omega_{0}^{-1}$ (blue dot-dashed line) and $T^{-1}=0.20\omega_{0}^{-1}$ (purple solid line). Other parameters are chosen as $\omega_{c}=5\omega_{0}$, $g_{0}=0.1\omega_{0}$, and $\omega_{0}=1$.}
\end{figure}
\begin{figure}
\centering
\includegraphics[angle=0,width=7cm]{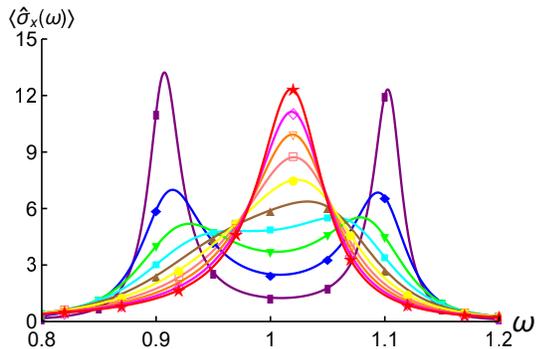}
\caption{\label{fig:fig4} The frequency spectrum $\langle\hat{\sigma}_{x}(\omega)\rangle$ of qubit $A$ versus $\omega$ with different coupling strengths: $\eta=0.001\omega_{0}$ (purple line with solid rectangles), $\eta=0.002\omega_{0}$ (blue line with solid diamonds), $\eta=0.003\omega_{0}$ (green line with solid down triangles), $\eta=0.004\omega_{0}$ (cyan line with solid squares), $\eta=0.005\omega_{0}$ (brown line with solid up triangles), $\eta=0.006\omega_{0}$ (yellow line with solid circles), $\eta=0.007\omega_{0}$ (pink line with open squares), $\eta=0.008\omega_{0}$ (orange line with open down triangles), $\eta=0.009\omega_{0}$ (magenta line with open diamonds) and $\eta=0.010\omega_{0}$ (red line with solid stars). Other parameters are chosen as $T^{-1}=0.03\omega_{0}^{-1}$, $\omega_{c}=5\omega_{0}$, $g_{0}=0.1\omega_{0}$, and $\omega_{0}=1$.}
\end{figure}
In this section, we would like to briefly discuss the physical reason for the occurrence of this anomalous deocherence phenomenon in this two-qubit spin-boson model. By making use of the generalized Silbey-Harris or Lang-Firsov transformation~\cite{7,8,9,10,11}, one can obtain the effective decoherence rate of qubit $A$ under Weisskopf-Wigner approximation as follows~\cite{11}
\begin{equation}\label{eq:eq21}
\gamma_{0}=const\times J_{\rm eff}(\omega_{d}),
\end{equation}
where $\omega_{d}$ denotes the dominant frequency of $\langle\hat{\sigma}_{x}(t)\rangle$ and $J_{\rm eff}(\omega)$ can be regarded as an effective bath spectral density function for qubit $A$ which is given by
\begin{equation*}
J_{\rm eff}(\omega)=\frac{1}{\pi}\frac{g_{0}^{2}\vartheta(|\omega|)}{[|\omega|-\zeta\omega_{0}-R(|\omega|)]^{2}+\vartheta^{2}(|\omega|)},
\end{equation*}
\begin{equation*}
R(\omega)=\wp\int_{0}^{\infty}d\omega'\frac{(\zeta\omega_{0})^{2}}{(\omega-\omega')(\omega'+\zeta\omega_{0})^{2}}J(\omega')\coth(\frac{\omega'}{2T}),
\end{equation*}
\begin{equation*}
\vartheta(\omega)=\pi(\frac{\zeta\omega_{0}}{\omega+\zeta\omega_{0}})^{2}J(\omega)\coth(\frac{\omega}{2T}),
\end{equation*}
where $\wp$ stands for the Cauchy principal value and $\zeta$ satisfies the self-consistent equation
\begin{equation*}
\zeta=\exp[-\frac{1}{2}\int_{0}^{\infty}d\omega \frac{J(\omega)}{(\zeta\omega_{0}+\omega)^{2}}\coth(\frac{\omega}{2T})].
\end{equation*}
From Eq.~\ref{eq:eq21}, one can find that the decoherence dynamics of qubit $A$ is determined by the dominant frequency $\omega=\omega_{d}$. In order to get insight into the dominant frequencies of $\langle\hat{\sigma}_{x}(t)\rangle$, a Fourier cosine transform is applied to $\langle\hat{\sigma}_{x}(t)\rangle$ according to
\begin{equation}\label{eq:eq22}
\langle\hat{\sigma}_{x}(\omega)\rangle\equiv\sqrt{\frac{2}{\pi}}\int_{0}^{\infty}dt\cos(\omega t)\langle\hat{\sigma}_{x}(t)\rangle.
\end{equation}
Then the frequency property of the dynamics $\langle\hat{\sigma}_{x}(t)\rangle$ can be analyzed directly by $\langle\hat{\sigma}_{x}(\omega)\rangle$~\cite{9,10,11,43}.

In Figs.~\ref{fig:fig3}(a) and (c), we display the frequency spectrum $\langle\hat{\sigma}_{x}(\omega)\rangle$ as the function of $\omega$ for strong coupling and weak coupling, respectively. We find that there is only one dominant frequency $\omega_{d1}\simeq\omega_{0}$ in the strong coupling regimes (see Fig.~\ref{fig:fig3}(a)). While, in weak coupling regimes, two characteristic frequencies $\omega_{d2,d3}\simeq\omega_{0}\pm g_{0}$ are dominating the quantum beat dynamics (see Fig.~\ref{fig:fig3}(c)), which is consistent with the well-known quantum beat phenomenon~\cite{11,43}. Thus in strong coupling regimes, $\gamma_{0}\propto J_{\rm eff}(\omega_{0})$, we find that the value of $J_{\rm eff}(\omega_{0})$ is reduced by increasing the bath temperature (see Fig.~\ref{fig:fig3}(b)), this is the reason of emergence of the anomalous decoherence phenomenon. However, in weak coupling case, it is $J_{\rm eff}(\omega_{0}\pm g_{0})$, rather than $J_{\rm eff}(\omega_{0})$ that determines the decoherence rate $\gamma_{0}$. In this situation, the value of $J_{\rm eff}(\omega_{0}\pm g_{0})$ is indeed enhanced with the increase of bath temperature. The dominant frequency shift from $\omega_{d1}\simeq\omega_{0}$ to $\omega_{d2,d3}\simeq\omega_{0}\pm g_{0}$ plays an important role in this situation and the effect of frequency shift cannot be predicted by using the Lindblad formula; thus the results of weak coupling were not reported in Ref.~\cite{30}.

We display the frequency spectrum $\langle\hat{\sigma}_{x}(\omega)\rangle$ for different coupling strength $\eta$ in Fig.~\ref{fig:fig4}. It can be seen from Fig.~\ref{fig:fig4} that the frequency spectrum $\langle\hat{\sigma}_{x}(\omega)\rangle$ transforms from a double-peak structure to a single peak with the increase of coupling strength $\eta$. In this sense, the critical coupling strength $\eta_{c}$, which determines the transition from quantum beat dynamics to damped oscillation, can be characterized by the behavior of the frequency spectrum $\langle\hat{\sigma}_{x}(\omega)\rangle$. It is interesting to point out that the frequency spectrum of the population difference of the spin-boson model exhibits a similar behavior close to the coherence-incoherence transition point~\cite{45} at zero temperature~\cite{9,10}.

The existence of qubit $B$ significantly changes the characteristics of the original bath spectral density function, i.e., $J(\omega)\rightarrow J_{\rm eff}(\omega)$. The engineered bath spectral density function $J_{\rm eff}(\omega)$ is responsible for the decoherence behaviors of the two-qubit spin-boson system and results in this anomalous decoherence phenomenon. In this sense, the decoherence behavior of a quantum dissipative system can be modulated by adding an assisted degree of freedom (qubit $B$ in this model). A similar scheme to modulate the decoherence behaviors of quantum dissipative system has also been reported in several previous studies~\cite{add3,add4}.

In summary, we study the decoherence of a two-qubit spin-boson model in the framework of stochastic decoupling along with HEOM without the usual Born-Markov or weak coupling approximations. It is shown that the decoherence of qubit $A$ can be reduced by increasing the bath temperature in strong-coupling regimes, which is contrary to the common recognition that a higher bosonic bath temperature always induces a more severe decoherence. For the weak coupling case, the quantum coherence of qubit $A$ is not meliorated but rather destroyed with the increase of bath temperature. These result suggest that there exists a decoherence dynamics transition point $\eta_{c}$ separating these two different decoherence behaviors. And we also show that the critical coupling strength $\eta_{c}$ can be characterized by the behavior of the frequency spectrum $\langle\hat{\sigma}_{x}(\omega)\rangle$. Finally, due to the generality of the qubit-oscillator model, we expect our results to be of interest for a wide range of experimental applications in quantum computation and quantum information processing.

\section{Acknowledgments}\label{sec:sec5}

W. Wu wishes to thank Dr. D.-W. Luo, Prof. H. Zheng, Prof. H.-G. Luo and Prof. G.-W. Wang for many useful discussions. W. Wu and H.-Q. Lin acknowledge the support from NSAF U1530401 and computational resources from the Beijing Computational Science Research Center.

\section{Appendix}\label{sec:sec6}

\begin{widetext}

In this Appendix, we show how to obtain the master equation of Eq.~\ref{eq:eq19} in the main text. The general non-Markovian master equation for a reduced density matrix $\tilde{\rho}_{s}(t)$ in the interaction picture is given by~\cite{39,40,41,42}
\begin{equation}\label{eq:eq23}
\frac{d}{dt}\tilde{\rho}_{s}^{I}(t)=-\int_{0}^{t}d\tau tr_{b}[\hat{H}_{sb}(t)^{\times}\hat{H}_{sb}(\tau)^{\times}\tilde{\rho}_{s}^{I}(\tau)\otimes\hat{\rho}_{th}],
\end{equation}
where we have made use of the Born approximation, i.e., we have assumed that the system and bath remain in the product state $\tilde{\rho}_{s}(t)\otimes\hat{\rho}_{th}$ for all the time. Operators are $\tilde{\rho}_{s}^{I}(t)\equiv e^{i\hat{H}_{s}t}\tilde{\rho}_{s}(t)e^{-i\hat{H}_{s}t}$, $\hat{\rho}_{th}\equiv \frac{e^{-\hat{H}_{b}/T}}{tr_{b}(e^{-\hat{H}_{b}/T})}$ and
\begin{equation}\label{eq:eq24}
\begin{split}
\hat{H}_{sb}(t)\equiv&e^{i(\hat{H}_{s}+\hat{H}_{b})t}\hat{H}_{sb}e^{-i(\hat{H}_{s}+\hat{H}_{b})t}\\
=&[e^{i\hat{H}_{s}t}f(\hat{s})e^{-i\hat{H}_{s}t}]\otimes [e^{i\hat{H}_{b}t}g(\hat{b})e^{-i\hat{H}_{b}t}]\\
=&\hat{f}_{s}(t)\otimes \hat{g}_{b}(t).
\end{split}
\end{equation}
Substituting Eq.~\ref{eq:eq24} into Eq.~\ref{eq:eq23} , the master equation can be rewritten as follows
\begin{equation}\label{eq:eq25}
\begin{split}
\frac{d}{dt}\tilde{\rho}_{s}^{I}(t)=&-\int_{0}^{t}d\tau [C_{R}(\tau)\hat{f}_{s}(t)^{\times}\hat{f}_{s}(t-\tau)^{\times}+iC_{I}(\tau)\hat{f}_{s}(t)^{\times}\hat{f}_{s}(t-\tau)^{\circ}]\tilde{\rho}_{s}^{I}(t-\tau).
\end{split}
\end{equation}
In the Markovian approximation, one can replace $\tilde{\rho}_{s}^{I}(t-\tau)$ by $\tilde{\rho}_{s}^{I}(t)$ and extend the upper limit of integral from $t$ to $+\infty$~\cite{39,40,41,42}. Then the Born-Markov master equation in the Schrodinger picture is given by
\begin{equation}\label{eq:eq26}
\begin{split}
\frac{d}{dt}\tilde{\rho}_{s}(t)\simeq&-i\hat{H}_{s}^{\times}\tilde{\rho}_{s}(t)-\int_{0}^{\infty}d\tau [C_{R}(\tau)\hat{f}_{s}^{\times}\hat{f}_{s}(-\tau)^{\times}+iC_{I}(\tau)\hat{f}_{s}^{\times}\hat{f}_{s}(-\tau)^{\circ}]\tilde{\rho}_{s}(t).
\end{split}
\end{equation}

In order to deal with the time-dependent term $\hat{f}_{s}(-\tau)$, we need to diagonalize the quantum subsystem $\hat{H}_{s}$ numerically. Let $|\varphi_{r}\rangle$ be the eigenstate of $\hat{H}_{s}$, i.e., $\hat{H}_{s}|\varphi_{r}\rangle=\varepsilon_{r}|\varphi_{r}\rangle$ with $r=1,2,3,4$. Then we can reexpress the operator $\hat{f}_{s}(t)$ in the eigenbasis $\{|\varphi_{1}\rangle,|\varphi_{2}\rangle,|\varphi_{3}\rangle,|\varphi_{4}\rangle\}$ as follows:
\begin{equation}\label{eq:eq27}
\begin{split}
\hat{f}_{s}(t)\equiv &e^{i\hat{H}_{s}t}f(\hat{s})e^{-i\hat{H}_{s}t}\\
=&\sum_{r,r'}\langle\varphi_{r}|f(\hat{s})|\varphi_{r'}\rangle e^{i(\varepsilon_{r}-\varepsilon_{r'})t}|\varphi_{r}\rangle\langle\varphi_{r'}|\\
=&\sum_{r,r'}f_{rr'}e^{i\varepsilon_{rr'}t}|\varphi_{r}\rangle\langle\varphi_{r'}|
\end{split}
\end{equation}
where $\varepsilon_{rr'}\equiv\varepsilon_{r}-\varepsilon_{r'}$ is the difference between the $r$th and $r'$th eigenvalues, $f_{rr'}\equiv\langle\varphi_{r}|f(\hat{s})|\varphi_{r'}\rangle$ denotes the jump matrix between eigenstates $|\varphi_{r}\rangle$ and $|\varphi_{r'}\rangle$. Using this definition, we can further simplify the operator $\Upsilon(\hat{s})$ as follows:
\begin{equation*}
\begin{split}
\Upsilon(\hat{s})\equiv&\int_{0}^{\infty}d\tau C_{R}(\tau)\hat{f}_{s}(-\tau)\\
=&\int_{0}^{\infty}\int_{0}^{\infty}d\tau d\omega J(\omega)\coth(\frac{\omega}{2T})\cos(\omega \tau)\sum_{r,r'}f_{rr'}e^{-i\varepsilon_{rr'}\tau}|\varphi_{r}\rangle\langle\varphi_{r'}|\\
=&\sum_{r,r'}\int_{0}^{\infty}d\tau\cos(\omega \tau)e^{-i\varepsilon_{rr'}\tau}\int_{0}^{\infty} d\omega J(\omega)\coth(\frac{\omega}{2T})f_{rr'}|\varphi_{r}\rangle\langle\varphi_{r'}|\\
=&\frac{1}{2}\sum_{r,r'}\int_{0}^{\infty}d\tau [e^{-i(\omega+\varepsilon_{rr'})\tau}+e^{i(\omega-\varepsilon_{rr'})\tau}]\int_{0}^{\infty} d\omega J(\omega)\coth(\frac{\omega}{2T})f_{rr'}|\varphi_{r}\rangle\langle\varphi_{r'}|.
\end{split}
\end{equation*}
We make use of the formula
\begin{equation*}
\int_{0}^{\infty}d\tau e^{\pm i\omega\tau}\simeq \pi\delta(\omega)\mp i\wp\frac{1}{\omega},
\end{equation*}
where $\delta(x)$ is the famous Dirac $\delta$ function. By neglecting the imaginary Lamb-shift terms~\cite{41,42}, we can obtain the approximate expression of $\Upsilon(\hat{s})$ as follows:
\begin{equation}\label{eq:eq28}
\begin{split}
\Upsilon(\hat{s})\simeq&\frac{\pi}{2}\sum_{r,r'}\int_{0}^{\infty} d\omega [\delta(\omega+\varepsilon_{rr'})+\delta(\omega-\varepsilon_{rr'})]J(\omega)\coth(\frac{\omega}{2T})f_{rr'}|\varphi_{r}\rangle\langle\varphi_{r'}|\\
=&\pi\sum_{r,r'}J(\varepsilon_{rr'})\coth(\frac{\varepsilon_{rr'}}{2T})f_{rr'}|\varphi_{r}\rangle\langle\varphi_{r'}|.
\end{split}
\end{equation}

Making use of the same method outlined above, one can also obtain
\begin{equation}\label{eq:eq29}
\begin{split}
\Xi(\hat{s})&\equiv-i\int_{0}^{\infty}d\tau C_{I}(\tau)f_{s}(-\tau)\simeq\pi\sum_{r,r}J(\varepsilon_{rr'})f_{rr'}|\varphi_{r}\rangle\langle\varphi_{r'}|.
\end{split}
\end{equation}
Finally, we obtain the Born-Markov master equation as follows
\begin{equation*}
\frac{d}{dt}\tilde{\rho}_{s}(t)=[-i\hat{H}_{s}^{\times}-f(\hat{s})^{\times}\Upsilon(\hat{s})^{\times}+f(\hat{s})^{\times}\Xi(\hat{s})^{\circ}]\tilde{\rho}_{s}(t),
\end{equation*}
which is Eq.~\ref{eq:eq19} in the main text.

\end{widetext}

\end{document}